%% file: lsc.tex
\newcommand{\CLASSINPUTtoptextmargin}{0.73in}   

\documentclass[conference]{IEEEtran}
\usepackage[numbers, square, comma, sort&compress]{natbib}

\ifCLASSOPTIONcompsoc
\else
\fi

\ifCLASSINFOpdf
\else
\fi

\usepackage[sf,SF]{subfigure}
\usepackage{psfrag}
\usepackage{pstricks}
\usepackage{epsfig}
\usepackage{amsmath,amsthm}
\usepackage{amssymb, bm, dsfont}
\usepackage{float}
\usepackage{amssymb,stmaryrd,amsmath,amsfonts,rotating}
\usepackage{multirow}
\usepackage{booktabs}

\providecommand{\setR}{\ensuremath{\mathbb{R}}}

\providecommand{\E}{\ensuremath{\mathbb{E}}}

\newcommand{\pr}{\mathbb{P}}

\newcommand{\X}{\mathcal{X}}
\newcommand{\U}{\mathcal{U}}

\newcommand{\abs}[1]{\left|#1\right|}

\newcommand{\expec}[2]{\ensuremath{\E_{#1}\left(#2\right)}}

\theoremstyle{definition}
\floatstyle{ruled} 
\newfloat{algorithm}{tbp}{loa}
\floatname{algorithm}{Algorithm}

\begin{document}
%
\title{Lossy Source Coding via Spatially Coupled LDGM Ensembles}
\author{
	\IEEEauthorblockN{
		Vahid Aref, Nicolas Macris, R{\"u}diger Urbanke and Marc Vuffray \\
		\IEEEauthorblockA{LTHC, IC, EPFL, Lausanne, Switzerland, 
		}}
}
\maketitle

\IEEEcompsoctitleabstractindextext{%
\begin{abstract}
We study a new encoding scheme for lossy source compression based
on spatially coupled low-density generator-matrix codes. 
We develop a belief-propagation 
guided-decimation algorithm, and show that this algorithm 
allows to  approach the optimal distortion of 
spatially coupled ensembles. 
Moreover, using the survey propagation formalism, we also observe that the optimal distortions of the spatially coupled and individual code ensembles are the same.
Since regular low-density generator-matrix codes are known to achieve the Shannon rate-distortion bound 
under optimal encoding as the degrees grow, our results suggest that spatial coupling can be used to reach the rate-distortion bound, under a {\it low complexity} belief-propagation 
guided-decimation algorithm. 

This problem is analogous to the MAX-XORSAT problem in computer science.  




\end{abstract}

}


\IEEEdisplaynotcompsoctitleabstractindextext

%
\IEEEpeerreviewmaketitle

\section{Introduction}

%
%

%
%
%
%

\IEEEPARstart{T}{he} spatial coupling of copies of a graphical code was introduced in \cite{ZigFel} in the form of convolutional low-density parity-check (LDPC) codes.
The performance of such ensembles under the belief propagation (BP) algorithm is consistently better than the performance of the underlying ensembles 
\citep{IDLDPCC,TerminLDPCCCthreshold,ProtoLDPCC}. 
The key observation is that the  BP threshold of a coupled ensemble considerably improves and gets close to the maximum a posteriori (MAP) threshold of the underlying ensemble.
This {\em threshold saturation} phenomenon has been  studied rigorously in \citep{CouplLDPC09,Coupl11BMS}.
Furthermore, it has been investigated in other models such as the Curie-Weiss chain, random satisfiability, graph coloring~\citep{Hassani10Couplgraphical,HMU10,Hassani11SAT}, and 
compressed sensing~\citep{Kudekar10compress,Krzakala10compress}.

One of the classic problems in communications is lossy source compression.
The objective is to compress a given sequence, so that it can be reconstructed up to some specified distortion. For binary symmetric sources,
low-density generator-matrix (LDGM) codes are able to asymptotically achieve Shannon's rate-distortion bound under optimal encoding (minimum distance
encoding) \citep{Wainwright10LDGManalysis,Ciliberti05PSC}. However, this is not a computationally efficient scheme. 

LDGM codes are well-suited to low-complexity message passing algorithms such as the BP algorithm. 
But using LDGM codes with a plain BP algorithm is not very effective in
lossy compression.  
To achieve more promising results, one can equip the BP algorithm with a decimation process. The general idea of belief-propagation guided-decimation (BPGD) algorithms is to: i) compute BP marginals, 
ii) fix bits with the largest bias, iii) decimate the graph. Decimation reduces the graph to a smaller one, on which this process is repeated.
Many variants of message passing algorithms
and various decimation processes have been investigated.
In \cite{Wainwirht05LDGMSP} a survey propagation (SP) inspired decimation algorithm is proposed. Simulations show 
distortions close to the rate-distortion bound 
for large block-lengths. 
Later, similar results were reported with
modified forms of BP  \citep{Filler07BPLDGM,CG10LSC}.
The performance of these algorithms also depends on the choice of degree distributions for LDGM codes. A heuristic choice is to use degree distributions that have been 
optimized for LDPC codes on binary symmetric channel \citep{Wainwirht05LDGMSP,Filler07BPLDGM,CG10LSC}. No rigorous analysis exists to date which can explain why this is the case.

In the present contribution, we study a spatially coupled LDGM ensemble for lossy
source compression. Two of us \cite{Aref11UniRateless} studied such ensembles in 
the context of rateless codes (channel coding) and demonstrated that threshold saturation takes place.
We provide
numerical evidence showing that a similar effect occurs in lossy compression. 
In particular, the BPGD distortion of the coupled LDGM ensemble
approaches (numerically) the optimal distortion of the underlying ensemble. 
We use the simplest 
forms of BP and decimation processes, and we take regular degrees. It is noteworthy that no 
optimization is needed, thus suggesting that the observed saturation is related to fundamental principles.
Regular (underlying) LDGM ensembles with large degrees have an optimal distortion that approaches the rate-distortion limit. 
Thus with spatially coupled LDGM codes with regular large degrees and a BPGD algorithm we can attain the Shannon limit. 
The complexity of the encoding scheme presented here is $O(n^2L^2)$ where $L$ is the number of copies of the underlying ensemble and $n$ the size 
of each copy. 

In section \ref{sec:framework},
we briefly review lossy compression and explain the structure of coupled LDGM ensembles. In section \ref{sec:optimal},
we formulate lossy compression as an optimization problem and compute the optimal distortion for underlying and coupled LDGM ensembles, by using the SP formalism. 
In section \ref{sec:BP},
we formulate and discuss the BPGD algorithm. 
Simulation results for this algorithm are presented in section \ref{sec:simulation}. 
A few practical issues are discussed in section \ref{sec:conclusion}.

\section{Framework}\label{sec:framework}

\subsubsection{Lossy Compression of Symmetric Bernoulli Sources}
Let $\underline{X}\in \X=\{0,1\}^n$ represent a binary source of length $n$. We
have $\underline{X} = \{X_1,X_2,\dots,X_n\}$, where $\{X_a\}_{a=1}^n$
are i.i.d random variables with
$\pr\{X_a = 1\} = \frac{1}{2}, \text{ for }a\in\{1,\dots,n\}$.
We compress a given source word $\underline{x}$  by mapping it to one of the $2^{nR}$ index words $\underline{u}\in\U = \{0,1\}^{nR}$, where $R \in [0, 1]$ is the rate. 
The stored sequence $\underline{u}$
 then defines a reconstructed source sequence
  $\underline{\widehat{x}}(\underline{u}) \in \X$, where the
   source decoding map $\underline{u}\to\underline{\widehat{x}}(\underline{u})$
 depends on the structure of the code.
For a given pair $(\underline{x},\underline{\widehat{x}})$,
 the distortion is measured
by Hamming distance $\frac{1}{n} d_H(\underline{x},\underline{\widehat{x}}) = \frac{1}{n}\sum_{a=1}^n \abs{x_a - \widehat{x}_a}$.
 Thus, the average quality of the reconstruction is
  measured by $D:=\frac{1}{n}\expec{\underline{X}}{d_H(\underline{x},\underline{\widehat{x}})}$.

For the symmetric Bernoulli source, it is
well-known that for any scheme, the average distortion is lower-bounded by
 $D_{\text{sh}}$, defined implicitly by the rate-distortion function
  $h(D_{\text{sh}}) = 1 - R,\; D_{\text{sh}}\in[0,\frac{1}{2}]$. Here $h(\cdot)$ is
  the binary entropy function. The goal is to find an encoding scheme such that, for a given $R$, it achieves the rate-distortion lower bound.

For this purpose, we study LDGM codes since they are able to achieve 
the rate-distortion bound under the optimal encoding when the average degrees increase \cite{Wainwright10LDGManalysis}.

\subsubsection{LDGM Codes}
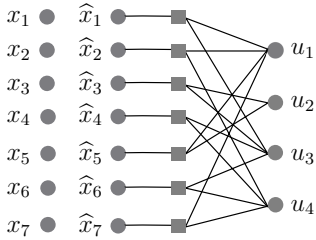
\begin{figure}[tb]
\setlength{\unitlength}{1.0bp}%
\centering
\scalebox{0.8}{\input{fig/ldgm}}
\caption{\label{fig:factorgraph} The factor graph associated to an underlying LDGM code.}
\end{figure}

Consider an LDGM code of block length $n$ and rate $R=\frac{m}{n}$.
Let $u_1,\cdots,u_m$ be $m$ code-bits of the index word $\underline{u}$
and let  $\widehat{x}_1,\widehat{x}_2,\cdots,\widehat{x}_n$ be $n$ reconstruction bits of the source word $x_1, x_2,\cdots, x_n$.
An LDGM code is usually represented by a bipartite graph as depicted in Figure~\ref{fig:factorgraph}. Call this graph 
$\Gamma(C,G;E)$. Each code-bit $u_i$ is represented by a node $i \in C(\Gamma)$. For each reconstruction bit $\widehat{x}_a$, there is a \emph{generator node} $a \in G(\Gamma)$. Each edge $(i,a)\in E(\Gamma)$ 
shows that the code-bit $u_i$ is connected to
the corresponding reconstruction bit $\widehat{x}_a$.
We denote by $\partial a$ the set of all code-bit nodes connected to $a\in G$, i.e.
$\partial a=\left\{  i\in C|\left(  i,a\right)  \in E\right\}  $. Similarly,
for $i\in C$, $\partial i=\left\{  a\in G|\left(  i,a\right)  \in E\right\}$. Thus, the reconstructed bit $\widehat{x}_{a}$ is obtained from the code-bits $\underline{u}$ by a  mod 2 sum
$\widehat{x}_{a}={\oplus_{i\in\partial a}} u_{i}$.
 
In this paper, we focus on the $(l, r,n)$-regular LDGM random ensemble, where 
$l$ (resp. $r$) is the degree of generator (resp. code-bit) nodes. We refer to \cite{URbMCT}
for the detailed  construction of this ensemble.

\subsubsection{Chain of LDGM$(l,r,n)$ Ensembles}

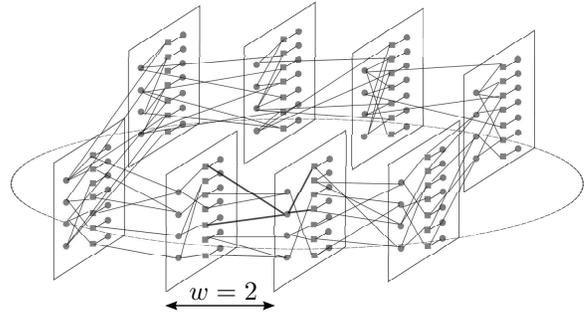
\begin{figure}[tb]
\setlength{\unitlength}{1.0bp}%
\centering
\input{fig/chain}
\caption{\label{fig:chain} Illustration of a factor graph from a CCLDGM ensemble with length $L=8$ and $w=2$.}
\end{figure}

Consider $L$ sets of nodes each having $m=\frac{l}{r}n$ code-bit nodes and
$n$ generator nodes (and reconstruction nodes).
Locate the sets in positions $0$ to $L-1$ on a circle
(see Figure~\ref{fig:chain}).
We randomly 
connect each generator node in position $z\in[0,L-1]$, to $l$ code-bit nodes from 
$w$ sets in the range $[z,z+w-1]$. Eventually (for large $n$ and $m$) code-bit nodes
have degree $r$.  The details of this construction are 
explained in \cite{CouplLDPC09,Aref11UniRateless}. 
Note that this ensemble has the same {\it local} structure as the underlying LDGM$(l,r)$ ensemble. 
Because of the {\it global} circular structure, we call it a \emph{Closed Chain LDGM$(l,r,L,w,n)$ ensemble} (CCLDGM).

We wish to point out a difference between the present construction and the ones used so far
in channel coding. The latter are based on open linear chains with suitable boundary conditions
that help initiate the decoding process, which then propagates like a wave through the system. In the present periodic construction, encoding will be seeded by preferential decimation at a specific position (say $z=L/2$), in an initial phase of the BPGD algorithm.
\section{Optimal Encoding and Distortion}\label{sec:optimal}

 
\subsection{Optimal Encoding}

%
Let $\Gamma$ be the factor graph of a random instance of an LDGM$(l,r,n)$ ensemble or CCLDGM$(l,r,L,w,n)$ ensemble.
We are looking for
a configuration $\underline{u}^{\ast}$ that minimizes the
distortion between $\underline{x}$ and $\widehat{\underline{x}}(\underline{u})$, i.e.%
\begin{equation}
d_{\min}:=\min_{\underline{u}}\frac{1}{nL}d_{H}\left( \underline{x},\widehat{\underline{x}}\right)  =\frac{1}{nL}d_H\left(\underline{x},
{\displaystyle\oplus_{i\in\partial a}}
u^{\ast}_{i}\right). 
\label{equ_argminu}%
\end{equation}
Note that for the individual ensemble $L=1$.
As we are
interested in the average distortion of the ensemble over all $\underline{X}\in \X$,
we define the optimal distortion as 
\begin{equation}
D_{\text{opt}} = \E_{\Gamma,\underline{X}}(d_{\text{min}}).
\end{equation}

We transform the above minimization problem into a maximum likelihood
problem by equipping the configuration space $\left\{  0,1\right\}
^{nLl/r}$ with the following probability measure,
\begin{equation}
\begin{aligned}
\mu_{\beta}(\underline{u}\mid \underline{x})
&:=\frac{1}{Z(\beta\mid\underline{x})} e^{-\beta d_H(\underline{x},\widehat{\underline{x}})}\\
&=\frac{1}{Z(\beta\mid\underline{x})}\prod_{a\in C(\Gamma)}%
e^{-\beta \vert x_a - {\bigoplus_{i\in\partial a}}u_i \vert},
\end{aligned}
\label{measure mu}%
\end{equation}
where $\beta\in \setR^{+}$ is a non-negative parameter, 
and $Z(\beta\vert\underline{x})  
=\sum_{\underline{u}}  e^{-\beta d_H(\underline{x},\widehat{\underline{x}})}$ the normalizing factor.
The minimizer $\underline{u}^{\ast}$ in
\eqref{equ_argminu} maximizes the measure, i.e, $\max_{\underline{u}}\mu_{\beta}\left(  \underline{u}\mid\underline{x}\right)
=\mu_{\beta}\left(  \underline{u}^{\ast}\mid\underline{x}\right)
$. Moreover, for a particular $\underline{x}$, the minimal distortion can be obtained as
$
d_{\min} = -(nL)^{-1}\lim_{\beta\rightarrow\infty}\beta^{-1}\ln Z\left(\beta\mid\underline{x}\right).
$
This formulation of the problem allows to use techniques from statistical physics to compute 
$D_{\text{opt}}$. In the physics interpretation, $d_H$ and $\mu_\beta$ are a random hamiltonian and Gibbs measure, $\beta$ is an inverse temperature, $Z$ a partition function, $D_{\text{opt}}$ is the average ground state energy. The latter can be computed by the SP formalism.
\subsection{Computation of the Optimal Distortion}
The details of the SP equations, used to compute  $D_{\text{opt}}$, are not shown here. A pedagogical account of the formalism can be found in \cite{mezard09information}. The numerical solution of the SP equations shows that the optimal distortions for the underlying LDGM$(l,r)$ and CCLDGM$(l,r,L,w)$ are {\it the same for each finite 
$w$ and $L$} (here $n$ and $m$ are infinite). In fact, for an ensemble with even $l$ and Poisson degree
for code-bit nodes, a rigorous proof is 
presented in \cite{Hassani11SAT} where the spatially coupled periodic XORSAT problem is discussed.
Optimal distortions for degree distributions $(k,2k)$, $k = 3,4\text{ and }5$ are given in the first line of Table \ref{table:static}.
We observe that, as $k$ increases, the optimal distortion converges to the 
rate-distortion bound $D_{\text{sh}}$. This observation is consistent with the result in \citep{Wainwright10LDGManalysis,Ciliberti05PSC}.

\begin{table}[tb]
\caption{The optimal and the BPGD distortions
for the LDGM$(k,2k)$ ensembles. The BPGD distortions are evaluated for $n=200000$ nodes. The rate-distortion bound for
$R=0.5$ is $D_{\text{sh}} \approx 0.1100$.}
\centering
{\input{fig/tablestatic}}
\label{table:static} 
\end{table}
\section{Belief Propagation Guided Decimation}\label{sec:BP}

\subsection{Belief Propagation Approximation}

The naive way to compute the partition function involves the summation over $2^{nLl/r}$ terms. 
Unfortunately, this approach
is too complex. The same problem occurs for computing the marginal distribution
of a node. 
On a tree, however, these computations can be done exactly and yield the BP equations. We deal with graphs that are locally tree-like and this motivates the use of the BP equations as a first approximation for the marginals.
These involve a set of $2\left\vert E\left(  \Gamma\right)
\right\vert $ real valued messages, each pair being associated to an edge. The messages on the edge $\left(  i,a\right)  \in C\left(  \Gamma\right)  \times G\left(
\Gamma\right)  \,$\ are denoted by $\eta_{i\rightarrow a}$ and $\widehat{\eta}_{a\rightarrow i}$, and satisfy
\begin{align}
\widehat{\eta}_{a\rightarrow i}  &  =\frac{1}{\beta}\tanh^{-1}\left(
(-1)^{x_{a}}\tanh(\frac{\beta}{2})  \prod_{j\in\partial a\backslash i}\tanh\left(
\beta\eta_{j\rightarrow a}\right)  \right)\nonumber\\
\eta_{i\rightarrow a}  &  =\sum_{b\in\partial i\backslash a}\widehat{\eta
}_{b\rightarrow i}.\label{Bp equations}
\end{align}
As a side remark, note that as $\beta\rightarrow\infty$ the BP equations become the so-called Min-Sum equations.
From the solutions of equations \eqref{Bp equations} one can compute 
the BP marginal distributions of code-bit $i$ and generator node $a$, and a Bethe approximation $\mu_\beta^{\text{BP}}(\underline u\vert \underline x)$  (this is not a probability measure in general)
of the 
original measure $\mu_{\beta}(\underline u\vert \underline x)$. We refer to \cite{mezard09information} for more information.

Unfortunately, the number of solutions of the BP equations which lead to roughly the same 
distortion, grows exponentially large in terms of $n$, and one cannot find the relevant 
fixed point by a plain iterative method.  
To resolve this problem, 
the BP iterations are equipped with a heuristic decimation 
process. This forms the basis of the BPGD algorithm.
\subsection{BPGD algorithm}

In this section we consider an instance of CCLDGM $(l,r,L,w,n)$ ensemble. Equations \eqref{Bp equations} will be solved iteratively starting from the initial conditions $\eta_{i\rightarrow a}^{\left(  0\right)
}=\widehat{\eta}_{a\rightarrow i}^{\left(  0\right)  }=0$. At iteration $t$ the messages are $\eta_{i\rightarrow a}^{\left(  t\right)}$ and $\widehat{\eta}_{a\rightarrow i}^{\left(  t\right) }$. We define the bias of a code-bit $i$ at time $t$ as
\begin{equation}
b_t\left(  i\right)  =\beta \sum_{a\in\partial i}\widehat{\eta}_{a\rightarrow
i}^{\left(  t\right)  }. \label{bias}
\end{equation}
This represents the tendency of the code-bit to be $0$ or $1$. Our BPGD algorithm 
shown in Algorithm~\ref{BPGD}, uses a {\it decimation condition}. Let $\epsilon>0$ a small quantity, $\alpha>0$ a large quantity and $T$ an iteration time. Typical values used in the simulations are 
$\epsilon=0.01$, $\alpha=4.25$ and $T=10$.
We say that the {\it decimation condition is fulfilled} if one of following occurs:

\begin{itemize}
\item[i)]
After some time $\tau(\epsilon)<T$, the messages do not change significantly in two successive iterations, in the sense that
 $ \frac{1}{nL} \sum_{(i,a) \in E(\Gamma)}\vert \widehat{\eta}_{a\rightarrow i}^{\left(  t \right)} - \widehat{\eta}_{a\rightarrow i}^{\left(  t-1\right)} \vert < \epsilon$ for $t>\tau(\epsilon)$.
\item[ii)]
For some $i$ and $t<T$, $\vert b_t(i)\vert> \alpha$. 
\item[iii)]
None of the above conditions has taken place  $t\leq T$. 
\end{itemize}

\begin{algorithm}[htb]
\caption{BPGD Algorithm}\label{BPGD}
\begin{enumerate}
\item Begin with the graph instance $\Gamma^{(0)}\in$ CCLDGM $(l,r,L,w,n)$.
\item $t=0$, Update equations \eqref{Bp equations} until the first time 
$t_1$ such that the decimation condition is fulfilled.
\item Find $B=\max_{i}\left\vert b_{t_1}\left(  i\right)  \right\vert$.
\item \textbf{If} $B = 0$, \textbf{then} randomly pick a code-bit $i$ from range $[\frac{L-w}{2},\frac{L+w}{2}]$ and fix it randomly to $0$ or $1$.  
\item[] \textbf{Else} pick a code-bit $i\in \left\{  j\in C\left(
\Gamma^{\left(  k\right)  }\right) \mid \vert b_{t_1}\left(  j\right)  \vert=B
\right\} $ randomly and fix 
$u_{i}=\frac{1}{2}(\text{sign}(b_{t_1}(i))+1)$.
\item Update $x_a^{(k+1)}=x_a^{(k)}-u_i$, for $a\in\partial i$, otherwise $x_a^{(k+1)}=x_a^{(k)}$.
Then, update $\Gamma^{(k+1)}=\Gamma^{\left(  k\right)  }\backslash\{i\}$.
\item \textbf{If} there exists an unfixed code-bit, \textbf{ then} go to (2), 
\item[] \textbf{Else} finish and return $\underline{u}$.
\end{enumerate}
\end{algorithm}

It is not difficult to see that the complexity of the algorithm is $O(n^2L^2)$. Note that after each decimation, rather than resetting messages to zero we continue with the previous messages. 


An important element in the algorithm is that when there is no significant bias (step (4)) the random decimation
occurs in a specific bounded interval centered around $L/2$ and of size $w$. We observe that 
this creates a seed from which the encoding process starts propagating through the ring. 
The usual hard decimation algorithms randomly choose (when there is no bias) a code-bit
from the whole graph. We observed that when this prescription is used for CCLDGM instances, 
the distortion does not improve with respect to that of usual LDGM instances. 
\subsection{Optimal $\beta$ and Relation to BSC}
We have seen that a configuration
$\underline{u}^{\ast}$ maximizes the probability $\mu_{\beta}\left(
\underline{u}\right)  $ for all $\beta>0$.  If $\mu_{\beta}^{\text{BP}}$ was an accurate
approximation of $\mu_{\beta}$, we would expect the maximum of $\mu_{\beta}^{\text{BP}}(  \underline{u})$ to be independent of $\beta$. But $\mu_{\beta}^{\text{BP}}$ is not
an exact description of $\mu_{\beta}$. It is then natural to look at the
performance of the BPGD procedure for different $\beta$ and find an optimal parameter $\beta_{\text{opt}}$.

\begin{figure}[tb]%
\centering
\includegraphics[
trim=0.000000in 0.000000in -0.000500in 0.000000in,
height=1.65in,
width=3.0268in
]%
{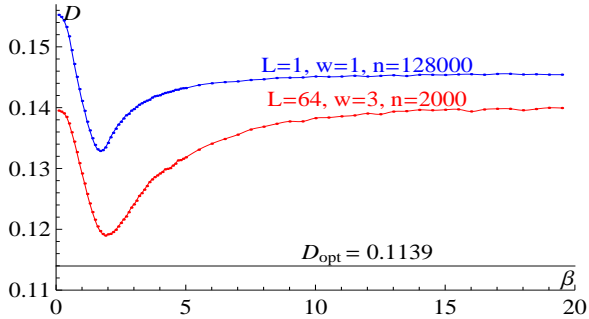}%
\caption{BPGD distortion versus $\beta$ for CCLDGM$(3,6,64,3,2000)$ (bottom) and LDGM$(3,6,128000)$ (top) ensembles.}%
\label{fig_optimal_beta}%
\end{figure}

Figure \ref{fig_optimal_beta} illustrates the performance of BPGD versus $\beta$ for CCLDGM$(3,6,64,3,2000)$ and LDGM$(3,6,128000)$  ensembles. For this example, the first
observation is that, for all $\beta>0$, the coupled ensemble has smaller $D_{\text{BPGD}}$ than the
uncoupled ensemble. We also observe that each ensemble has an optimal
parameter $\beta_\text{opt}$ which lies between $\frac{3}{2}$ and $\frac{5}{2}$. In order to make comparison between coupled and uncoupled ensembles and as the distortion
does not vary much when $\beta\in\left[\frac{3}{2},\frac{5}{2}\right]$, we fix $\beta=2$
 in the rest of the paper. Table~\ref{table:static} shows that 
 the BPGD distortion - for the value $\beta=2$ - is still consistently 
 higher than $D_{\text{opt}}$. 

The measure $\mu_{\beta}$ also arises in the context of channel
coding over a memoryless binary symmetric channel (BSC). Suppose that we use the LDGM code with factor graph $\Gamma$ over a BSC with flipping probability $p$. Then the posterior probability that the un-coded
message $\underline{u}\in\mathbb{F}_{2}^{nl/r}$ is sent given the received
codeword $\underline{x}\in\mathbb{F}_{2}^{n}$ can be formally expressed just as in \eqref{measure mu}
with $\beta \to \ln\left(  \frac{1-p}{p}\right)$. 
The difference between channel coding and lossy source
coding lies in the distribution of $\underline{X}$. In lossy source coding, $\{X_{a}\}$
are $n$ i.i.d. Ber$(1/2)$ variables; whereas in
channel coding over a BSC, they are not in general
i.i.d. and depend on the flipping probability $p$ (or $\beta$).

\section{Distortion Saturation in a CCLDGM Ensemble}\label{sec:simulation}

\begin{figure*}[htb]
\setlength{\unitlength}{0.8bp}%
\centering
\input{fig/constellation}
\caption{\label{fig:constellation} BPGD distortion profiles for a
periodic chain of $(5,10,L,w,n)$ 
CCLDGM ensembles. The values of $(L,w)$ are shown on the plots and $n=2000$. 
} 
\end{figure*}
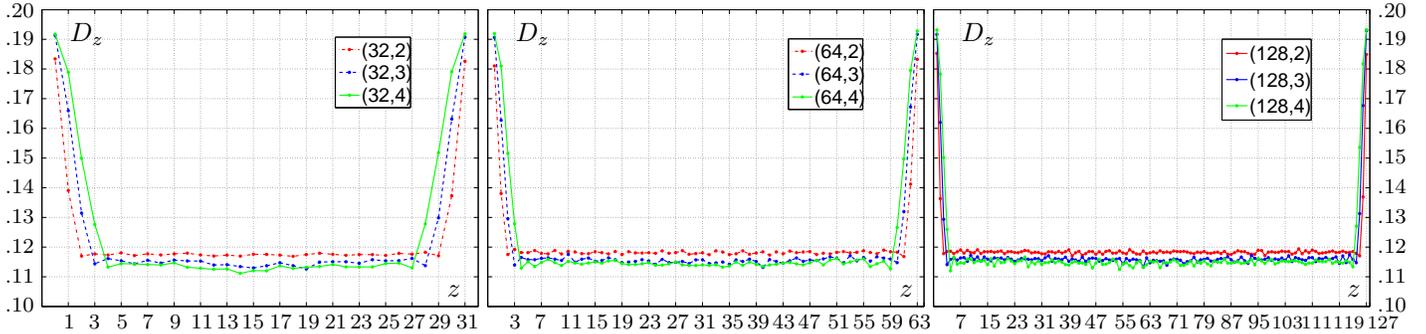

We provide simulation results which convincingly show that the BP distortion of a CCLDGM ensemble closely approaches the optimal distortion of the underlying LDGM ensemble as $n$, $L$ and $w$ grow large.
To emphasize the effect of coupling,
we consider the family of $(k,2k)$-regular LDGM ensembles. This family is known
to yield a weak performance under the BPGD algorithm~\cite{Wainwright10LDGManalysis}. Indeed, although the
optimal distortion is rather close to the rate-distortion bound and improves as $k$ grows,
the BPGD distortion becomes larger as $k$ grows (see Table~\ref{table:static}).
Analogous behaviors are well known to occur in channel coding~\cite{URbMCT}.

Now consider a CCLDGM$(k,2k,L,w)$ ensemble. Let $D_z$ denote the average distortion of the reconstruction nodes lying at position $z$. Call $D_z$ the \emph{local distortion at position $z$}. Thus, the total average distortion is $D = \frac{1}{L}\sum_{z=0}^{L-1} D_z$. Call the vector $(D_0,D_1,\cdots,D_{L-1})$ the \emph{distortion profile}.

When we apply the BPGD algorithm, we observe a generic behaviour in the distortion profile. Figure~\ref{fig:constellation} illustrates
the distortion profile of CCLDGM$(5,10,L,w,n)$ ensembles for different pairs of $(L,w)$ and  $n=2000$ . The profile is symmetric around $\frac{L}{2}$ (this expected in view of the algorithm).
Consider the first $\frac{L}{2}$ components. In the range $[0, w-1]$, the local distortion is strictly decreasing. It starts at a value roughly equal to $D_{\text{BPGD}}$ of the underlying ensemble and decreases to a value that we call the \emph{saturation value}.
Then, in the range $[w, \frac{L}{2}]$, the local distortions nearly remain constant and equal to the saturation value.
We refer to the first interval and the second interval respectively as \emph{the unsaturated part} and \emph{the saturated part}. Therefore, the average
distortion considerably decreases in the coupled ensembles.


Tables \ref{table:table36}, \ref{table:table48} and \ref{table:table510} show the saturation value (in bold) and the average distortion (in parentheses) for $k=3,4, 5$ and  different values of $n,L$ and $w$. Each value is averaged over 100 random 
code instances and random source words.
Inspection of the tables suggests that it approaches $D_{\text{opt}}$ in the 
regime $n\gg L\gg w\gg 1$. The average
 distortion $D_{\text{BPGD}}$ converges to the saturation value by increasing $L$. The reason is the unsaturated part essentially does not change for fixed $k, w$ and $n$ (see Figure~\ref{fig:constellation}), thus its contribution in the average distortion vanishes in the large $L$ limit. As a result, we expect that $D_{\text{BPGD}}$ gets very close to $D_{\text{opt}}$ for a large enough $n\gg L\gg w\gg 1$ and the optimal $\beta$. Moreover, if $k$ grows, $D_{\text{opt}}$ converges to the rate-distortion bound. Thus, the $D_{\text{BPGD}}$ of a regular CCLDGM ensemble can get close to the rate-distortion bound. 
\begin{table}[tb]
\centering
\caption{\label{table:table36} Saturated (bold)
and average BPGD distortion (parentheses) for CCLDGM$(3,6,L,w,n)$.}
\renewcommand{\arraystretch}{1.3}
\scalebox{1}{\input{fig/table36}}
\end{table}

\begin{table}[tb]
\centering
\renewcommand{\arraystretch}{1.3}
\caption{\label{table:table48} Saturated (bold) and average BPGD distortion (parentheses) for CCLDGM$(4,8,L,w,n)$.}
\scalebox{1}{\input{fig/table48}}
\end{table}
 \begin{table}[tb]
\centering
\caption{\label{table:table510} Saturated (bold) and average BPGD distortion (parentheses) for
 CCLDGM$(5,10,L,w,n)$.}
\renewcommand{\arraystretch}{1.3}
\scalebox{1}{\input{fig/table510}}
\end{table}


As mentioned before, the coupled chains used in channel coding are terminated by suitable boundary conditions and this incurs a rate loss of order 
$O(\frac{w}{L})$. In CCLDGM ensembles,
we do not pay this
cost in the graph structure, but
it manifests itself in the large local distortion in the  $2w$ positions of the unsaturated part.
 

\section{Conclusion}\label{sec:conclusion}
We have observed that CCLDGM$(k,2k,L,w)$ ensembles
can asymptotically saturate the rate-distortion bound under a BPGD algorithm. Degrees $(k,2k)$ have been 
chosen because it is known they lead to weaker results in the uncoupled case, but
the saturation presumably occurs for a large class of irregular LDGM ensembles.
As a result, using a right-regular CCLDGM ensemble
with Poisson distribution on the
code-bit nodes, would allow to
achieve the rate-distortion bound for {\it all} rates $R\in(0,1)$.

There are clearly a number of features of the algorithm that can be improved. Two important issues are its convergence rate,
and its complexity.
To address the first one, we can fix to zero
the code-bits lying in range
$[\frac{(L-w)}{2}, \frac{L+w}{2}]$ and then, we remove the reconstruction 
nodes (and source nodes) lying in ranges $[0,\frac{w}{2}]$ and $[L-1-\frac{w}{2},L-1]$. The total rate of 
the code slightly increases but the convergence rate is improved.
The complexity of the BPGD algorithm used here is $O(n^2L^2)$. 
To reduce it one could use a ``sliding window
 encoding'' similar to the one used for coupled LDPC codes~\cite{TerminLDPCCCthreshold}. This would reduce
 the complexity to  $O(n^2L)$. 
Finally the use of soft decimation techniques might also help improve the overall performance.



\vspace{0.8 cm}
\bibliographystyle{IEEEtran}
\footnotesize
\bibliography{references}

\appendices

\end{document}

%% file: fig/ldgm.tex
\begin{picture}(130,105)(0,0)
{

\put(0,0){\rotatebox{0}{\includegraphics[scale=0.7]{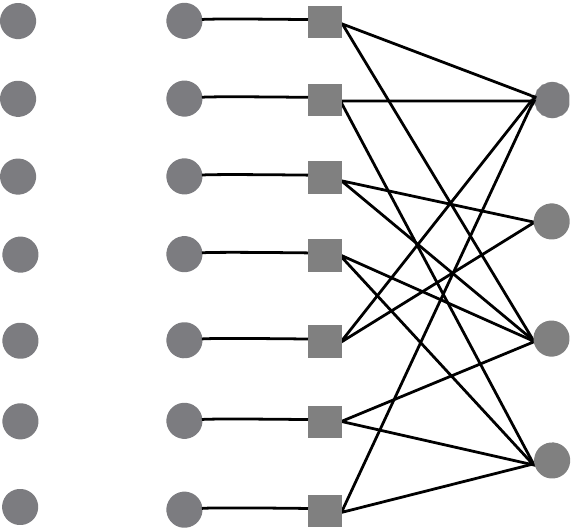}}}
\large{
\put(130,9){\makebox(0,0)[rb]{$u_4$}}
\put(130,33){\makebox(0,0)[rb]{$u_3$}}
\put(130,58){\makebox(0,0)[rb]{$u_2$}}
\put(130,83){\makebox(0,0)[rb]{$u_1$}}

\put(30,98){\makebox(0,0)[rb]{$\widehat{x}_1$}}
\put(30,82){\makebox(0,0)[rb]{$\widehat{x}_2$}}
\put(30,66){\makebox(0,0)[rb]{$\widehat{x}_3$}}
\put(30,51){\makebox(0,0)[rb]{$\widehat{x}_4$}}
\put(30,33){\makebox(0,0)[rb]{$\widehat{x}_5$}}
\put(30,17){\makebox(0,0)[rb]{$\widehat{x}_6$}}
\put(30,-1){\makebox(0,0)[rb]{$\widehat{x}_7$}}

\put(-4,98){\makebox(0,0)[rb]{${x}_1$}}
\put(-4,82){\makebox(0,0)[rb]{${x}_2$}}
\put(-4,66){\makebox(0,0)[rb]{${x}_3$}}
\put(-4,51){\makebox(0,0)[rb]{${x}_4$}}
\put(-4,33){\makebox(0,0)[rb]{${x}_5$}}
\put(-4,17){\makebox(0,0)[rb]{${x}_6$}}
\put(-4,-1){\makebox(0,0)[rb]{${x}_7$}}
}
}
\end{picture}

%% file: fig/chain.tex
\begin{picture}(210,110)(0,0)
{
\put(0,0){\rotatebox{0}{\includegraphics[scale=0.55]{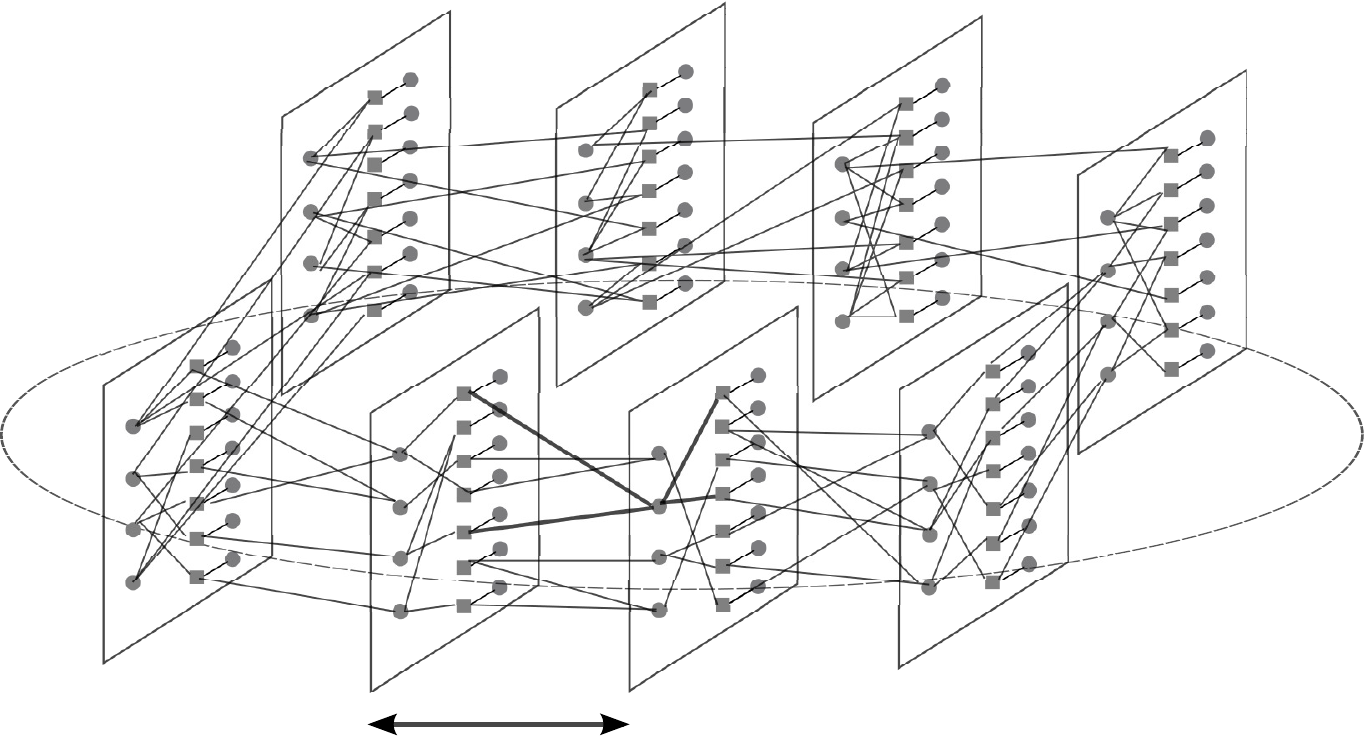}}}
\put(93,4){\makebox(0,0)[rb]{$w=2$}}
}
\end{picture}

%% file: fig/tablestatic.tex

\begin{tabular}{@{}llll@{}}\toprule
$(k,2k)$ & $(3,6)$  & $(4,8)$ & $(5,10)$\\
\midrule
$D_{\text{opt}}$ &  0.1139 & 0.1111 &  0.1105\\
$D_{\text{BPGD}}$ & 0.1357 & 0.1590 &   0.1811 \\
\bottomrule
\end{tabular}

%% file: fig/constellation.tex
\begin{picture}(630,140)(0,0)
{

\put(0,0){\rotatebox{0}{\includegraphics[scale=0.21]{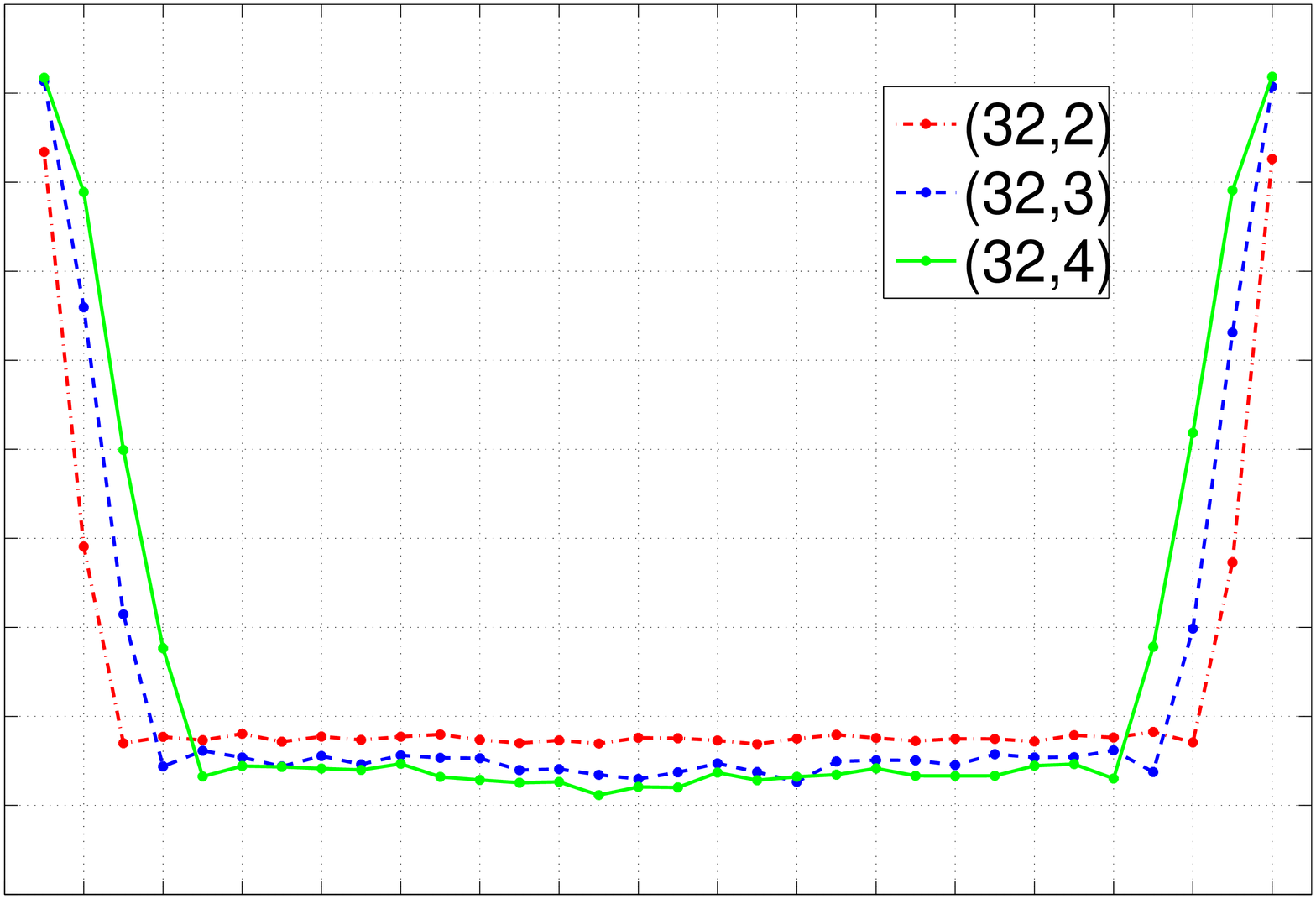}}}
\put(210,0){\rotatebox{0}{\includegraphics[scale=0.21]{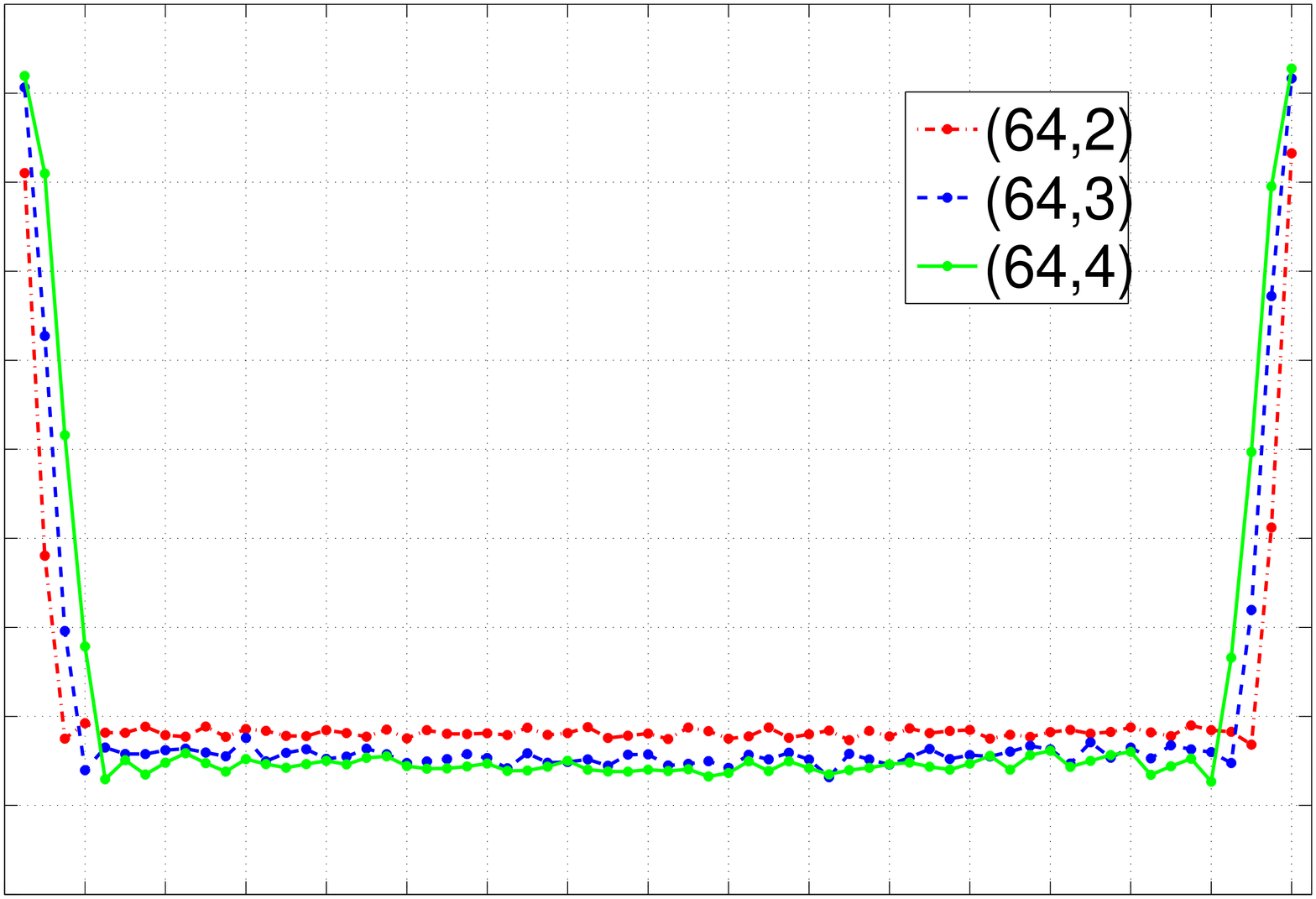}}}
\put(420,0){\rotatebox{0}{\includegraphics[scale=0.21]{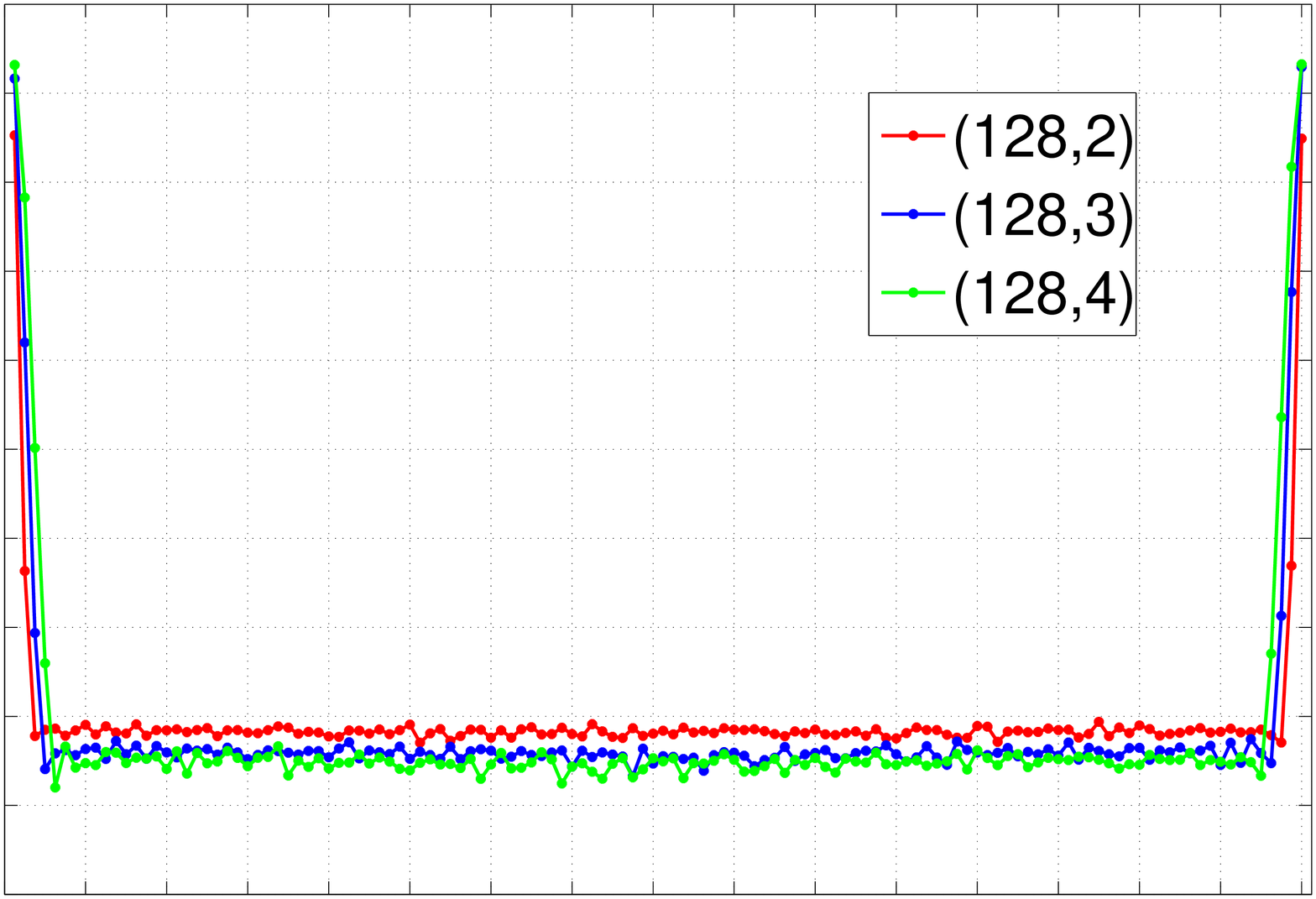}}}

\normalsize
{
\put(30,125){\makebox(0,0)[rb]{$D_z$}}
\put(240,125){\makebox(0,0)[rb]{$D_z$}}
\put(450,125){\makebox(0,0)[rb]{$D_z$}}
\put(198,6){\makebox(0,0)[rb]{$z$}}
\put(411,6){\makebox(0,0)[rb]{$z$}}
\put(622,6){\makebox(0,0)[rb]{$z$}}
}
\footnotesize
{
\put(-3,-1){\makebox(0,0)[rb]{$.10$}}
\put(-3,13){\makebox(0,0)[rb]{$.11$}}
\put(-3,27){\makebox(0,0)[rb]{$.12$}}
\put(-3,41){\makebox(0,0)[rb]{$.13$}}
\put(-3,55){\makebox(0,0)[rb]{$.14$}}
\put(-3,69){\makebox(0,0)[rb]{$.15$}}
\put(-3,83){\makebox(0,0)[rb]{$.16$}}
\put(-3,97){\makebox(0,0)[rb]{$.17$}}
\put(-3,111){\makebox(0,0)[rb]{$.18$}}
\put(-3,125){\makebox(0,0)[rb]{$.19$}}
\put(-3,138){\makebox(0,0)[rb]{$.20$}}
}
\footnotesize
{
\put(630,-1){\makebox(0,0)[lb]{$.10$}}
\put(630,13){\makebox(0,0)[lb]{$.11$}}
\put(630,27){\makebox(0,0)[lb]{$.12$}}
\put(630,41){\makebox(0,0)[lb]{$.13$}}
\put(630,55){\makebox(0,0)[lb]{$.14$}}
\put(630,69){\makebox(0,0)[lb]{$.15$}}
\put(630,83){\makebox(0,0)[lb]{$.16$}}
\put(630,97){\makebox(0,0)[lb]{$.17$}}
\put(630,111){\makebox(0,0)[lb]{$.18$}}
\put(630,125){\makebox(0,0)[lb]{$.19$}}
\put(630,138){\makebox(0,0)[lb]{$.20$}}
}
\footnotesize
{
\put(11,-9){\makebox(0,0)[lb]{$1$}}
\put(23,-9){\makebox(0,0)[lb]{$3$}}
\put(36,-9){\makebox(0,0)[lb]{$5$}}
\put(48,-9){\makebox(0,0)[lb]{$7$}}
\put(61,-9){\makebox(0,0)[lb]{$9$}}
\put(71,-9){\makebox(0,0)[lb]{$11$}}
\put(83,-9){\makebox(0,0)[lb]{$13$}}
\put(95,-9){\makebox(0,0)[lb]{$15$}}
\put(108,-9){\makebox(0,0)[lb]{$17$}}
\put(121,-9){\makebox(0,0)[lb]{$19$}}
\put(134,-9){\makebox(0,0)[lb]{$21$}}
\put(146,-9){\makebox(0,0)[lb]{$23$}}
\put(159,-9){\makebox(0,0)[lb]{$25$}}
\put(171,-9){\makebox(0,0)[lb]{$27$}}
\put(183,-9){\makebox(0,0)[lb]{$29$}}
\put(196,-9){\makebox(0,0)[lb]{$31$}}
}
\footnotesize
{
\put(221,-9){\makebox(0,0)[lb]{$3$}}
\put(233,-9){\makebox(0,0)[lb]{$7$}}
\put(245,-9){\makebox(0,0)[lb]{$11$}}
\put(257,-9){\makebox(0,0)[lb]{$15$}}
\put(269,-9){\makebox(0,0)[lb]{$19$}}
\put(282,-9){\makebox(0,0)[lb]{$23$}}
\put(295,-9){\makebox(0,0)[lb]{$27$}}
\put(308,-9){\makebox(0,0)[lb]{$31$}}
\put(321,-9){\makebox(0,0)[lb]{$35$}}
\put(333,-9){\makebox(0,0)[lb]{$39$}}
\put(345,-9){\makebox(0,0)[lb]{$43$}}
\put(357,-9){\makebox(0,0)[lb]{$47$}}
\put(371,-9){\makebox(0,0)[lb]{$51$}}
\put(383,-9){\makebox(0,0)[lb]{$55$}}
\put(396,-9){\makebox(0,0)[lb]{$59$}}
\put(409,-9){\makebox(0,0)[lb]{$63$}}
}
\footnotesize
{
\put(431,-9){\makebox(0,0)[lb]{$7$}}
\put(442,-9){\makebox(0,0)[lb]{$15$}}
\put(455,-9){\makebox(0,0)[lb]{$23$}}
\put(468,-9){\makebox(0,0)[lb]{$31$}}
\put(480,-9){\makebox(0,0)[lb]{$39$}}
\put(492,-9){\makebox(0,0)[lb]{$47$}}
\put(506,-9){\makebox(0,0)[lb]{$55$}}
\put(518,-9){\makebox(0,0)[lb]{$63$}}
\put(531,-9){\makebox(0,0)[lb]{$71$}}
\put(543,-9){\makebox(0,0)[lb]{$79$}}
\put(556,-9){\makebox(0,0)[lb]{$87$}}
\put(569,-9){\makebox(0,0)[lb]{$95$}}
}
\scriptsize
{
\put(580,-9){\makebox(0,0)[lb]{$103$}}
\put(594,-9){\makebox(0,0)[lb]{$111$}}
\put(609,-9){\makebox(0,0)[lb]{$119$}}
\put(625,-9){\makebox(0,0)[lb]{$127$}}
}

}
\end{picture}

%% file: fig/table36.tex
\begin{tabular}{@{}lllll@{}}\toprule

\multicolumn{1}{l}{{\multirow{2}{*}{$L$}}}& \multicolumn{1}{l}{{\multirow{2}{*}{$w$}}} & \multicolumn{3}{c}{$n$}\\ \cmidrule{3-5}
& & \multicolumn{1}{c}{2000} & \multicolumn{1}{c}{4000} & \multicolumn{1}{c}{8000}\\
\midrule

\multirow{2}{*}{32} & 2 & \textbf{0.1182} (0.1204) &\textbf{0.1170} (0.1191) & \textbf{0.1162} (0.1183)\\
& 3 & \textbf{0.1171} (0.1205) & \textbf{0.1161} (0.1195) & \textbf{0.1157} (0.1190)\\ 
\midrule
\multirow{2}{*}{64} & 2 & \textbf{0.1185} (0.1195)& \textbf{0.1172} (0.1183) & \textbf{0.1164}(0.1174)\\
 & 3 & \textbf{0.1175} (0.1191) & \textbf{0.1166} (0.1182) & \textbf{0.1159} (0.1175) \\
\midrule
\multirow{2}{*}{128} & 2 & \textbf{0.1186} (0.1191)  & \textbf{0.1173} (0.1179) &  \\
 & 3 & \textbf{0.1176} (0.1184) & \textbf{0.1167} (0.1175) &   \\

\bottomrule
\end{tabular}


%% file: fig/table48.tex
\begin{tabular}{@{}lllll@{}}\toprule

\multicolumn{1}{l}{{\multirow{2}{*}{$L$}}}& \multicolumn{1}{l}{{\multirow{2}{*}{$w$}}} & \multicolumn{3}{c}{$n$}\\ \cmidrule{3-5}
& & \multicolumn{1}{c}{2000} & \multicolumn{1}{c}{4000} & \multicolumn{1}{c}{8000}\\
\midrule

\multirow{3}{*}{32} 

& 2 & \textbf{0.1165} (0.1205) & \textbf{0.1148} (0.1188) & \textbf{0.1134} (0.1175) \\

& 3 & \textbf{0.1146} (0.1212) & \textbf{0.1133} (0.1199) & \textbf{0.1124} (0.1189) \\ 
& 4 & \textbf{0.1135} (0.1226) & \textbf{0.1125} (0.1215) & \textbf{0.1118} (0.1208)  \\ 
\midrule

\multirow{3}{*}{64} 
					& 2 & \textbf{0.1170} (0.1190) & \textbf{0.1151} (0.1171) & \textbf{0.1138} (0.1157)  \\
 					& 3 & \textbf{0.1153} (0.1186) & \textbf{0.1139} (0.1172) & \textbf{0.1130} (0.1161)    \\
 					& 4 & \textbf{0.1144} (0.1189) & \textbf{0.1133} (0.1177) & \textbf{0.1125} (0.1169)   \\ 
\midrule
\multirow{3}{*}{128} 
& 2 & \textbf{0.1172} (0.1182) & \textbf{0.1153} (0.1162) &  \\
 & 3 & \textbf{0.1156} (0.1172) & \textbf{0.1141} (0.1157) &   \\
 & 4 & \textbf{0.1148} (0.1170) & \textbf{0.1136} (0.1158) &   \\

\bottomrule
\end{tabular}

%

%% file: fig/table510.tex
\begin{tabular}{@{}llllll@{}}\toprule

\multicolumn{1}{l}{{\multirow{2}{*}{$L$}}}& \multicolumn{1}{l}{{\multirow{2}{*}{$w$}}} & \multicolumn{3}{c}{$n$}\\ \cmidrule{3-5}
& & \multicolumn{1}{c}{2000} & \multicolumn{1}{c}{4000} & \multicolumn{1}{c}{8000}\\
\midrule

\multirow{4}{*}{32}& 2 & \textbf{0.1175} (0.1229) & \textbf{0.1153} (0.1207) & \textbf{0.1138} (0.1192) \\
					& 3 & \textbf{0.1147} (0.1236) & \textbf{0.1130} (0.1219) &  \textbf{0.1120} (0.1208) \\
					& 4 & \textbf{0.1134} (0.1256) & \textbf{0.1120} (0.1242) &  \textbf{0.1111} (0.1234) \\
					& 5 & \textbf{0.1124} (0.1280) & \textbf{0.1112} (0.1268) &  \textbf{0.1107} (0.1262) \\
					\midrule
\multirow{4}{*}{64} & 2 & \textbf{0.1181} (0.1208) & \textbf{0.1158} (0.1185) & \textbf{0.1140} (0.1167)  \\
 					& 3 & \textbf{0.1155} (0.1199) &  \textbf{0.1138} (0.1182) & \textbf{0.1125} (0.1169)    \\
 					& 4 & \textbf{0.1144} (0.1205) & \textbf{0.1130} (0.1190) & \textbf{0.1119} (0.1179)   \\
					& 5 & \textbf{0.1137} (0.1213) & \textbf{0.1124} (0.1200) & \textbf{0.1116} (0.1192) \\ 
					\midrule
\multirow{4}{*}{128} & 2 & \textbf{0.1182} (0.1196) & \textbf{0.1158} (0.1172) &  \\
 					 & 3 & \textbf{0.1158} (0.1180) & \textbf{0.1140} (0.1161) &   \\
 					 & 4 & \textbf{0.1148} (0.1178) & \textbf{0.1133} (0.1162) &   \\
 					 & 5 & \textbf{0.1141} (0.1179) & \textbf{0.1128} (0.1166) &   \\
 					 \bottomrule

\end{tabular}

%

%% file: lsc.bbl
\newcommand{\SortNoop}[1]{}
\begin{thebibliography}{10}
\providecommand{\url}[1]{#1}
\csname url@samestyle\endcsname
\providecommand{\newblock}{\relax}
\providecommand{\bibinfo}[2]{#2}
\providecommand{\BIBentrySTDinterwordspacing}{\spaceskip=0pt\relax}
\providecommand{\BIBentryALTinterwordstretchfactor}{4}
\providecommand{\BIBentryALTinterwordspacing}{\spaceskip=\fontdimen2\font plus
\BIBentryALTinterwordstretchfactor\fontdimen3\font minus
  \fontdimen4\font\relax}
\providecommand{\BIBforeignlanguage}[2]{{%
\expandafter\ifx\csname l@#1\endcsname\relax
\typeout{** WARNING: IEEEtran.bst: No hyphenation pattern has been}%
\typeout{** loaded for the language `#1'. Using the pattern for}%
\typeout{** the default language instead.}%
\else
\language=\csname l@#1\endcsname
\fi
#2}}
\providecommand{\BIBdecl}{\relax}
\BIBdecl

\bibitem{ZigFel}
A.~J. Felstrom and K.~S. Zigangirov, ``{T}ime-varying periodic convolutional
  codes with low density parity check matrix,'' \emph{IEEE Transactions on
  Information Theory}, vol.~45, no.~5, pp. 2181--2190, 1999.

\bibitem{IDLDPCC}
M.~Lentmaier, A.~Sridharan, D.~J. Costello, Jr., and K.~S. Zigangirov,
  ``Iterative decoding threshold analysis for {L}{D}{P}{C} convolutional
  codes,'' \emph{To appear in \emph{ IEEE Transactions on Information Theory}},
  2008.

\bibitem{TerminLDPCCCthreshold}
M.~Lentmaier, A.~Sridharan, K.~S. Zigangirov, and D.~J. {Costello},
  ``{T}erminated {L}{D}{P}{C} convolutional codes with thresholds close to
  capacity,'' in \emph{ISIT}, 2005, pp. 1372--1376.

\bibitem{ProtoLDPCC}
M.~Lentmaier, D.~G.~M. Mitchell, G.~P. Fettweis, and D.~J. Costello,
  ``{A}symptotically regular {L}{D}{P}{C} codes with linear distance growth and
  thresholds close to capacity,'' in \emph{Information Theory and Applications
  Workshop (ITA)}, January 2010, pp. 1--8.

\bibitem{CouplLDPC09}
S.~Kudekar, T.~J. Richardson, and R.~L. Urbanke, ``Threshold saturation via
  spatial coupling: why convolutional {L}{D}{P}{C} ensembles perform so well
  over the {B}{E}{C},'' \emph{http://arxiv.org/abs/1001.1826}, 2009.

\bibitem{Coupl11BMS}
S.~{Kudekar}, T.~{Richardson}, and R.~{Urbanke}, ``{Spatially Coupled Ensembles
  Universally Achieve Capacity under Belief Propagation},'' \emph{ArXiv}, Jan.
  2012.

\bibitem{Hassani10Couplgraphical}
S.~H. Hassani, N.~Macris, and R.~L. Urbanke, ``Coupled graphical models and
  their thresholds,'' in \emph{Information Theory Workshop}, Dublin, 2010.

\bibitem{HMU10}
------, ``Chains of mean field models,'' \emph{submitted to J. Stat. Mech:
  Theory and Experiment}, 2010.

\bibitem{Hassani11SAT}
------, ``Threshold saturation in spatially coupled constraint satisfaction
  problems,'' \emph{ArXiv}, Dec 2011.

\bibitem{Kudekar10compress}
S.~Kudekar and H.~Pfister, ``The effect of spatial coupling on compressive
  sensing,'' in \emph{48th Annual Allerton Conference}, Aug. 2010, pp. 347
  --353.

\bibitem{Krzakala10compress}
F.~{Krzakala}, M.~{M{\'e}zard}, F.~{Sausset}, Y.~{Sun}, and L.~{Zdeborov{\'a}},
  ``{Statistical physics-based reconstruction in compressed sensing},''
  \emph{ArXiv}, Sep. 2011.

\bibitem{Wainwright10LDGManalysis}
M.~Wainwright, E.~Maneva, and E.~Martinian, ``Lossy source compression using
  low-density generator matrix codes: Analysis and algorithms,''
  \emph{Information Theory, IEEE Transactions on}, vol.~56, no.~3, pp. 1351
  --1368, March 2010.

\bibitem{Ciliberti05PSC}
S.~Ciliberti and M.~Mézard, ``The theoretical capacity of the parity source
  coder,'' \emph{Journal of Statistical Mechanics: Theory and Experiment}, vol.
  2005, no.~10, p. P10003, 2005.

\bibitem{Wainwirht05LDGMSP}
M.~Wainwright and E.~Maneva, ``Lossy source encoding via message-passing and
  decimation over generalized codewords of {LDGM} codes,'' in \emph{ISIT},
  2005, pp. 1493 --1497.

\bibitem{Filler07BPLDGM}
T.~Filler and J.~Fridrich, ``Binary quantization using belief propagation with
  decimation over factor graphs of ldgm codes,'' in \emph{Allerton}, 2007.

\bibitem{CG10LSC}
D.~Castanheira and A.~Gameiro, ``Lossy source coding using belief propagation
  and soft-decimation over ldgm codes,'' in \emph{PIMRC}, sept. 2010, pp. 431
  --436.

\bibitem{Aref11UniRateless}
V.~Aref and R.~L. Urbanke, ``Universal rateless codes from coupled lt codes,''
  in \emph{Information Theory Workshop (ITW)}, oct. 2011, pp. 277 --281.

\bibitem{URbMCT}
T.~Richardson and R.~Urbanke, \emph{Modern Coding Theory}.\hskip 1em plus 0.5em
  minus 0.4em\relax Cambridge University Press, 2008.

\bibitem{mezard09information}
M.~M{\'e}zard and A.~Montanari, \emph{Information, physics, and computation},
  ser. Oxford graduate texts.\hskip 1em plus 0.5em minus 0.4em\relax Oxford
  University Press, 2009.

\end{thebibliography}
